\documentclass{article}
\usepackage[preprint]{log_2022}

\usepackage{booktabs}           
\usepackage{multirow}           
\usepackage{amsmath, amsthm, amsfonts, dsfont}
\usepackage{graphicx}           
\usepackage{duckuments}         

\usepackage{bm}
\usepackage{lipsum}
\usepackage{float}
\usepackage{subfigure}
\usepackage{wrapfig}

\usepackage[numbers,compress,sort]{natbib}

\title[3D Molecular Geometry Analysis with 2D Graphs]{3D Molecular Geometry Analysis with 2D Graphs}

\author{
  Zhao Xu\thanks{Equal contribution.} \\
  Texas A\&M University \\
  College Station, TX 77843\\
  \texttt{zhaoxu@tamu.edu} \\
  \And
  Yaochen Xie\footnotemark[1]  \\
  Texas A\&M University \\
  College Station, TX 77843\\
  \texttt{ethanycx@tamu.edu} \\
  \And
  Youzhi Luo \\
  Texas A\&M University \\
  College Station, TX 77843\\
  \texttt{yzluo@tamu.edu} \\
  \And
  Xuan Zhang \\
  Texas A\&M University \\
  College Station, TX 77843\\
  \texttt{xuan.zhang@tamu.edu} \\
  \And
  Xinyi Xu \\
  Xidian University \\
  Xi'an, 710071, China \\
  \texttt{xyxu.xd@gmail.com} \\
  \And
  Meng Liu \\
  Texas A\&M University \\
  College Station, TX 77843\\
  \texttt{mengliu@tamu.edu} \\
  \And
  Kaleb Dickerson \\
  Texas A\&M University \\
  College Station, TX 77843\\
  \texttt{kaleb.dickerson2001@tamu.edu} \\
  \And
  Cheng Deng \\
  Xidian University \\
  Xi'an, 710071, China \\
  \texttt{chdengxd@gmail.com} \\
  \And
  Maho Nakata \\
  RIKEN \\
  Wako, Saitama 351-0198, Japan \\
  \texttt{maho@riken.jp} \\
  \And
  Shuiwang Ji \\
  Texas A\&M University \\
  College Station, TX 77843\\
  \texttt{sji@tamu.edu} \\  
}

\begin{document}

\maketitle

\begin{abstract}
Ground-state 3D geometries of molecules are essential for many molecular analysis tasks.
Modern quantum mechanical methods can compute accurate 3D geometries but are computationally prohibitive. 
Currently, an efficient alternative to computing ground-state 3D molecular geometries from 2D graphs is lacking. Here, we propose a novel deep learning framework to predict 3D geometries from molecular graphs. To this end, we develop an equilibrium message passing neural network (EMPNN) to better capture ground-state geometries from molecular graphs.
To provide a testbed for 3D molecular geometry analysis, we develop a benchmark that includes a large-scale molecular geometry dataset, data splits, and evaluation protocols.
Experimental results show that EMPNN can efficiently predict more accurate ground-state 3D geometries than RDKit and other deep learning methods.
Results also show that the proposed framework outperforms self-supervised learning methods on property prediction tasks.
\end{abstract}

\section{Introduction}
In recent years, significant progress has been made in data-driven molecular representation learning methods. It becomes popular in many methods \cite{yang2019analyzing,stokes2020deep,wang2020advanced} to represent the structural information in forms of SMILES \cite{weininger1988smiles} sequences and 2D molecular graphs. Although molecular graphs represent atoms/bonds as nodes/edges to show connections explicitly, the specific shape of a molecule still lacks in 2D graphs. Compared to molecular graphs, 3D molecular geometries can provide additional spatial information which benefits the prediction of related properties. A molecule may correspond to multiple spatial arrangements of atoms known as conformations. Generally, every conformation, or equivalently 3D molecular geometry, corresponds to a state of specific potential energy. In particular, the equilibrium ground-state geometry is achieved when a molecule has the lowest possible energy due to the minimized repulsion and maximized attraction. Existing studies~\cite{wu2018moleculenet, gilmer2017neural} have demonstrated that the use of ground-state 3D geometries as inputs of machine learning models leads to more accurate quantum property prediction than using 2D graphs only. This improvement is consistent with the fact that properties are often closely related to the shape of a molecule~\cite{engel2018applied}. 

Despite the importance of ground-state 3D geometries, they are challenging and expensive to obtain. Currently, quantum chemistry calculations such as Gaussian~\cite{frisch2005gaussian}, QChem~\cite{shao2015advances}, MolPro~\cite{werner2012molpro}, NWChem~\cite{valiev2010nwchem}, Turbomole~\cite{furche2014turbomole}, and GAMESS~\cite{schmidt1993general} are adopted as common methods for computing accurate 3D geometries of molecules in more economical manners. 
Even so, it is a time-consuming process. For example, obtaining the precise 3D geometries of a small molecule using density functional theory (DFT) requires hours of computations. Currently, the high computational cost seriously impedes the wide usage of accurate ground-state 3D geometries in molecular analysis and simulation.

A promising way to reduce the computational cost is to reconstruct ground-state 3D geometries from 2D molecular graphs using machine learning models. Currently, there exist some generative methods aiming to generate 3D molecular geometries. Nevertheless, generative methods consider one-to-many problems, thus producing multiple molecular geometries that are not necessarily in the ground-state. In this work, we propose to predict the most stable ground-state 3D geometry from 2D molecular graphs in a one-step and end-to-end fashion. The proposed 3D geometry prediction is a novel problem that is fundamentally different from the problem solved by generative methods. We believe our work may provide a new avenue for the fast computation of ground-state 3D molecular geometries.

Fast computation of molecular geometries is motivated by the desire of utilizing 3D information in a variety of applications, including but not limited to, analysis of the molecular dynamics, prediction of biological activities of molecules, and design ligands or 3D linkers \cite{wu2018moleculenet, gilmer2017neural} for proteins. However, due to the poor accuracy of current fast geometry generative methods, few prior works employ the generated 3D geometries for downstream applications. Here, we propose a two-staged Geometry-Aware Prediction (GAP) framework composed of the novel 3D geometry prediction problem and molecular property predictions based on predicted geometries. GAP leverages our newly developed equilibrium message passing neural network (EMPNN) for predicting ground-state 3D geometries and employs a 3D graph neural network (GNN) model for property prediction.

The key contributions of this work include: (1) we formulate a novel problem of computing ground-state 3D molecular geometries from 2D graphs using predictive deep learning methods, (2) we propose a 2-stage framework in which a novel EMPNN is developed for ground-state 3D geometry prediction and a 3D GNN is adopted to use predicted geometries on downstream property predictions, (3) we establish the first benchmark with a large AI-ready dataset named Molecule3D together with evaluation protocols to enable the 3D molecular geometry analysis. Experimental results show that compared with baseline methods, EMPNN achieves more accurate ground-state 3D geometries with faster computation. Furthermore, property prediction results on MoleculeNet datasets show that the proposed framework outperforms existing self-supervised learning methods, which demonstrates the value and significance of the 3D geometry prediction problem.

\textbf{Relations with Prior Work.} In contrast to 3D GNNs~\cite{schutt2017schnet, liu2021spherical, klicpera2020directional, klicpera2020fast} that use 3D molecular geometries as input, we aim to predict 3D geometries from molecular graphs, which is an upstream task to those 3D GNNs. In addition, unlike generative methods~\cite{simm2020generative, mansimov2019molecular, xu2021an, shi2021learning, xu2021learning, ganea2021geomol} that learn the distribution of geometries, we study the predictive ground-state geometry which is more closely associated with molecule properties. Hence, datasets commonly used by generative methods are inapplicable in our problem. To enable the geometry predictive task, we construct the Molecule3D dataset. Compared to the OGB-PCQM4Mv2~\cite{hu2021ogb} dataset, our Molecule3D includes more molecules with 3D geometries, more molecular properties, and we provide data splits and evaluation protocols specifically for the proposed task. Furthermore, we compare with SSL pretraining methods~\cite{you2020graph, you2021graph, liu2021pre} to show that the proposed geometry predictive task and the curated Molecule3D dataset are practically meaningful.

\section{Method}
\textbf{Notations.} We treat a 2D molecule as a directed graph $G=(V, E)$, where $V$ is a set of nodes denoting atom in the molecule, and $E$ is a set of edges denoting bonds between atoms. We denote $(u, v)\in E$ the edge from $u$ to $v$, $\bm h_v$ the node-level embedding vector and $\bm e_{(u,v)}$ the edge-level embedding vector. Given a node $v\in V$, we denote $\mathcal{N}(v)=\{u: (v, u)\in E\}$ the set of neighbor nodes of $v$.

\subsection{The Geometry-Aware Prediction Framework}
\label{sec: framework}
The ground-state 3D molecular geometries are usually critical to determine the physical, chemical, and biological behaviors of molecules. 
However, existing deep learning-based quantitative analyses of structure-activity and structure-property are generally studied independently to the generation of 3D molecular geometries.
Aiming at fully incorporating molecular geometries in the study of molecule properties, we propose the Geometry-Aware Prediction (GAP) framework in this work. GAP specifically studies the ground-state geometry of molecules and conducts a comprehensive analysis consisting of two stages; namely ground-state 3D geometry prediction from molecular graphs, and downstream property prediction based on predicted geometries. We describe the two stages as follows.

\textbf{3D Geometry Prediction.} In contrast to generative or auto-regressive methods, we propose to predict 3D geometry in a one-step and end-to-end fashion. The 3D geometry prediction problem can be formulated as predicting the 3D coordinates of all atoms in the molecule or predicting the intra-molecular pairwise distances between atoms. Whereas there exist infinite sets of correct 3D coordinates for a given conformation, the pairwise distances are unique due to invariance to translation and rotation of the conformation. On the other hand, the predicted pairwise distances can fail to reconstruct atom coordinates due to violations of triangle inequalities as demonstrated in Appendix~\ref{append: math_coords}, leading to invalid molecule geometries.
To circumvent this problem, we let the geometry prediction model $f$ predict 3-dimensional vectors representing atom coordinates $f(G)\in \mathbb{R}^{|V|\times 3}$ intermediately and supervised the model on the pairwise distances $D\in\mathbb{R}^{|V|\times |V|}$ computed from $f(G)$.
This design allows consistent prediction of valid atom coordinates in 3D space and invariant inference of the model regardless of translations or rotations on ground-truth conformations.

\textbf{Downstream Property Prediction.} At the second stage, the ground-state 3D geometries learned in the first stage are employed to assist molecular property prediction inductively on a separate dataset. Cooperating with 3D GNNs, the ground-state 3D geometries can be utilized in two schemes. First, one can directly use the predicted coordinates or distance as fixed inputs of 3D GNNs for property prediction. Second, one can also adopt the 3D geometry prediction as a pre-training task. The geometry prediction model can be fine-tuned together with 3D GNNs for property prediction on downstream tasks.

The GAP framework is anticipated to demonstrate that predictive deep learning methods are promising on the fast and accurate computation of 3D molecular geometries, thereby overcoming the dilemma of scarce 3D geometry data and improving structure-related downstream tasks such as property prediction. An overview of the GAP framework is shown in Figure~\ref{fig:framework} in Appendix~\ref{append: overview}. In the next subsection, we focus on the most challenging problem in the framework, the ground-state geometry prediction, and propose the equilibrium message passing neural networks.

\subsection{Equilibrium Message Passing Neural Networks}
\label{sec: agg_rule}
Molecule geometry modeling methods such as the VSEPR theory~\citep{gillespie2013vsepr} and the electron localization~\citep{savin1992electron} assume that the arrangement of valence electrons that form either chemical bonds or lone pairs always minimize the repulsion effects from one another to achieve equilibrium. When it comes to molecules containing multiple central atoms, the ground-state geometry of a molecule is achieved when the local state of every atom geometry, \textit{i.e.}, their electron arrangement, reach an equilibrium with their neighbors'. Based on this intuition, we consider geometric states of individual atoms and propose to update the states of geometry repeatedly according to the states of neighbor atoms to equilibrate their geometric states.

Unlike 3D GNN models that require roto-translation equivariance~\citep{fuchs2020se3transformers} (roto-equivariance), the prediction of the ground-state geometry requires roto-invariant models, as geometric states of atoms are expected to remain the same when rotating a molecule. However, existing GNNs that include edge embeddings to capture information about bond lengths and absolute directions fail to represent geometric states in a roto-invariant manner. GNNs that simply aggregate information of node embeddings can be roto-invariant, but fail to capture full geometric characteristics of atoms and hence are unable to equilibrate their geometric states. Therefore, we propose the roto-invariant atom geometry representations and the equilibrium message passing to describe and update states of atom geometries towards the equilibrium of all atom geometries.

\begin{wrapfigure}[16]{R}{6.5cm}
    \centering
    \vspace{-10pt}
    \includegraphics[width=0.46\textwidth]{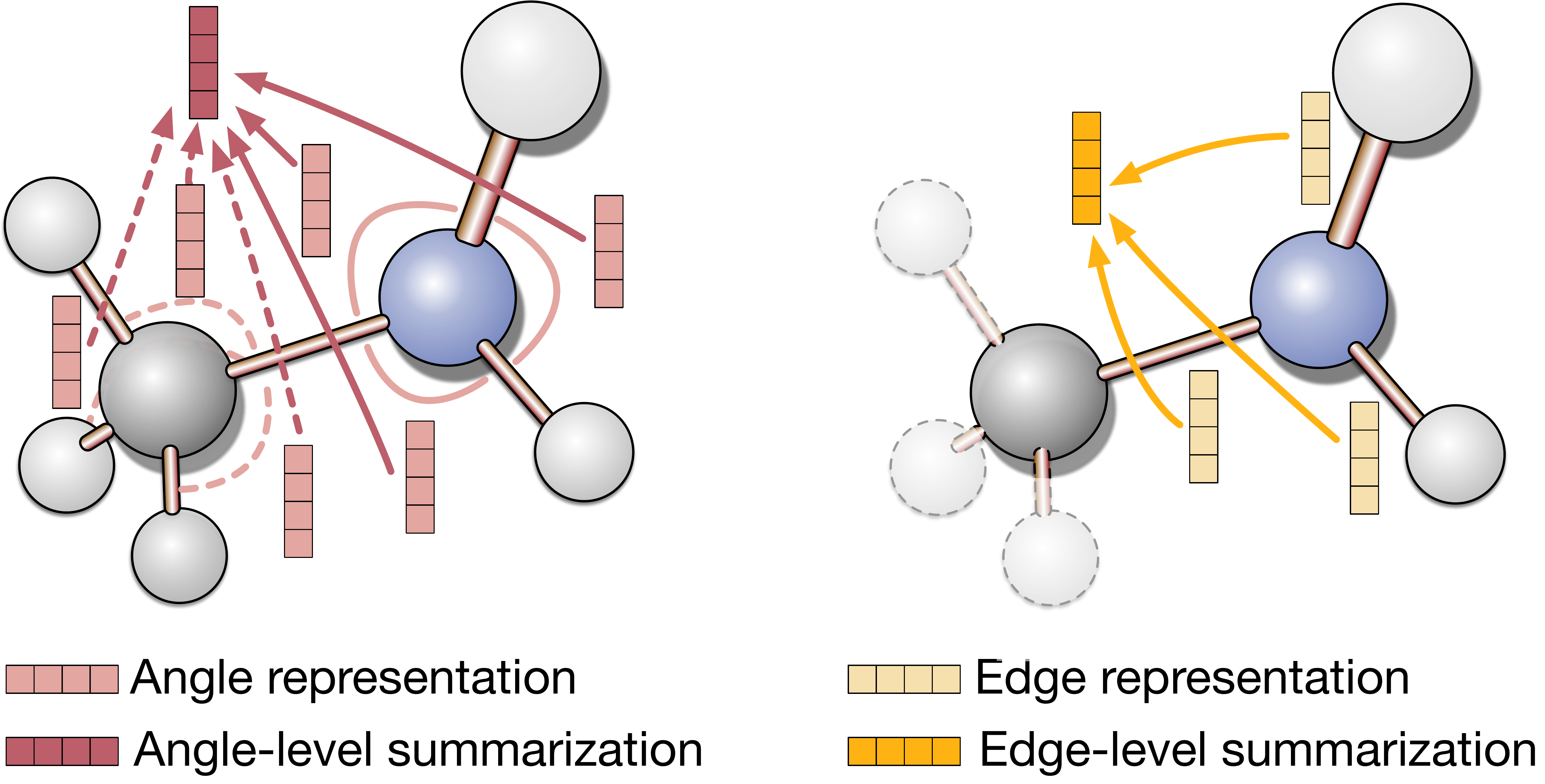}
    \vspace{-15pt}
    \caption{Summarizations of angle-level and edge-level representations. For angle-level summarization, solid and dashed arrows indicate different MLP projections. The state of the atom in blue are to be updated with the summarizations in the next step.}
    \label{fig:empnn}
\end{wrapfigure}

\textbf{Roto-Invariant Atom Geometry Representations}. 
We consider an atom $v$ in a molecular graph $G$. Instead of representing absolute directions of bonds, we describe the state of the atom geometry by specifying the lengths of its bonds $\{(v, u)|u\in\mathcal{N}(v)\}$, and the angle (relative direction) between each pair of bonds $\{(u_i, v, u_j)|u_i, u_j\in\mathcal{N}(v), i\ne j\}$, as they are invariant to any rotations of a molecule. Correspondingly, we embed the state of atom $v$ with two sets of hidden representations corresponding to individual bonds and bond pairs (angles), denoted as the set of edge-level representations $H_e(v) = \{\bm e_{(v, u)}|u\in\mathcal{N}(v)\}$ and the set of angle-level representations $H_r(v)=\{r_{(u_i, v, u_j)}|u_i, u_j\in\mathcal{N}(v), i\ne j\}$, respectively. 
We additionally include an atom embedding $\bm h_v$ to capture information about lone electron pairs (that are not described by bonds but are critical for atom geometries) and summarize the bond and angle embeddings during updating. The complete representation of an atom geometric state is hence denoted as a triplet $S(v)=\big(\bm h_v, H_e(v), H_r(v)\big)$. The representations can capture sufficient information to describe the state of an atomic geometry and factors that affect the state in a roto-translation invariant manner.

\textbf{Equilibrium Message Passing}.
The equilibrium message passing (EMP) updates the state $S(v)$ of an atom according to its current state and the geometric states of its neighbor atoms. EMP is formulated as
\begin{equation}
    S^{(i+1)}(v) = \mbox{EMP}\Big(S^{(i)}(v), \{S^{(i)}(u)|u\in\mathcal{N}(v)\}\Big),
\end{equation}
where $S^{(i)}$ denotes the current state of an atom and $S^{(i+1)}$ denotes the updated state computed by the EMP layer. We include superscripts $\cdot^{(i)}$ and $\cdot^{(i+1)}$ in later equations to indicate the current and the updated states. Concretely, the EMP computes the updating of geometric states, namely all elements in the triplet $S(v)$, in following steps.

\textit{Step 1}: to update edge-level representations by aggregating states of neighbor atom geometries. Considering the state of node $v$ and the representation $\bm e_{(v, u)}\in H_e(v)$ corresponding to an edge $(v, u)$, the update is formulated as
\begin{equation}
    \bm e^{(i+1)}_{(v, u)} = \mbox{MLP}_e\left(\left[\bm e^{(i)}_{(v, u)}; \bm h^{(i)}_v; \bm h^{(i)}_u\right]\right),
\end{equation}
where $[\cdot;\cdot;\cdot]$ denotes the concatenation of vectors and $\mbox{MLP}_e(\cdot)$ denotes a multilayer perceptron.

\textit{Step 2}: to update angle-level representations with individual pairs of updated edge representations. Considering the representation $\bm r_{(u_i, v, u_j)}\in h_r(v)$ corresponding to the angle between edges $(v, u_i)$ and $(v, u_j)$, the update is formulated as
\begin{equation}
    \bm r^{(i+1)}_{(u_i, v, u_j)} = \mbox{MLP}_r\left(\left[\bm r^{(i)}_{(u_i, v, u_j)}; \bm e^{(i+1)}_{(v, u_i)}; \bm e^{(i+1)}_{(v, u_j)}\right]\right).
\end{equation}
\textit{Step 3}: to summarize all updated angle-level and edge-level representations related to the node $v$, as illustrated in Figure~\ref{fig:empnn}. For the angle-level summarization, we consider both cases when node $v$ serves as a center node of an angle, \textit{e.g.}, $(u_i, v, u_j)$, or a side node of an angle, \textit{e.g.}, $(v, u_i, u_j)$. Concretely, we compute the summarization $\bm m^{(r\to n)}_v$ of angle-level representations by
\begin{equation}
\begin{gathered}
    \bm m^{(r\to n, i+1)}_v = \mbox{MLP}_{ac}\Big(\sum H^{(i+1)}_r(v)\Big)+\\
    \mbox{MLP}_{as}\Big(\sum\{\bm r_{(v, u_i, u_j)}: u_i\in\mathcal{N}(v), (u_j, u_i)\in E\}\\
    \cup \{\bm r_{(u_j, u_i, v)}: u_i\in\mathcal{N}(v), (u_i, u_j)\in E\}
    \Big),
\end{gathered}
\end{equation}
where $\sum\{\cdot\}$ denotes the summation of all representations in the set. Moreover, the edge-level summarization $\bm m^{(e\to n)}_v$ is computed as
\begin{equation}
\bm m^{(e\to n, i+1)}_v = \mbox{MLP}_{e\to n}\Big(\sum H^{(i+1)}_e(v)\Big).
\end{equation}
\textit{Step 4}: to update atom embeddings (node representations) with the summarizations. Given the node $v$, the representation is updated by
\begin{gather}
    \bm h^{(i+1)}_v = \mbox{MLP}_h\left(\left[\bm h^{(i)}_v; \bm m^{(e\to n, i+1)}_v; \bm m^{(r\to n, i+1)}_v\right]\right).
\end{gather}
Batch normalization is adopted to all updated representations. Finally, the network predicts atom coordinates $\boldsymbol{x}_v$ from the representation at the last layer with an MLP. The above aggregation rules allows the update of individual bond and angle states, which enables states of  all atom geometries to reach the equilibrium.

\textbf{Initial Representations of Atom Geometry.}
We embed into initial representations both chemical features required to infer the atom geometry and coarse geometric features that are cheap in terms of computational cost. For node-level representations, in addition to the atom type, chirality, charges, \textit{etc.} adopted in OGB~\citep{hu2020open}, we further include positional embeddings based on the unique order of atoms in the atoms block of molfile~\cite{dalby1992description} that is the common raw structural format of molecules. The positional embeddings distinguish all atoms to ensure that symmetric atoms do not share identical representations, enabling the end-to-end prediction of molecule geometry in contrast to the auto-regressive generation methods. For edge-level representations, we embed the bond type, conjugation, and stereo following OGB. For the angle-level representations, we compute coarse angles based on 2D coordinates swiftly generated by RDKit and embed the angles with the spherical harmonics~\citep{liu2021spherical, klicpera2021gemnet}, formulated as
\begin{equation}
    Y_{\ell0}(\theta)=\sqrt{\frac{2\ell+1}{4\pi}}\frac{1}{2^\ell\ell!}\Big(\frac{d}{d\cos\theta}\Big)^\ell\big(\cos^2\theta-1\big)^\ell,
\end{equation}
where $\theta$ is an angle in radians and $\ell\in\{1,2,3\}$ corresponds to the 3 embedding dimensions.

\section{Relations with Existing Methods}
While 3D geometries are generally useful in many molecular analysis tasks, obtaining such information is highly nontrivial. In this section, we explain the differences between the GAP framework and existing methods that aim to obtain or utilize 3D molecular geometries.

\textbf{Non-Machine Learning Methods.} Instead of extensive experiments, first-principle quantum chemistry calculations can compute highly-accurate 3D geometries but are still computationally prohibitive. Non-machine learning methods such as EDG, ETKDG~\cite{riniker2015better}, and DG+UFF~\cite{rappe1992uff} are relatively faster alternatives but have much lower accuracy. In contrast to these non-machine learning methods that use physicochemical rules like force fields or expert knowledge, the GAP framework leverages a predictive deep neural network to predict 3D geometries from molecular graphs. 

\textbf{Generative Methods.}
Recently, some studies~\cite{simm2020generative, mansimov2019molecular, xu2021an, shi2021learning, xu2021learning, ganea2021geomol} propose to efficiently generate multiple 3D geometries of a molecule with deep generative models. Generative methods aim to learn the underlying distribution of the data, thus they typically consider one-to-many problems, i.e., generating multiple non-ground-state molecular geometries of the given molecular graph. On the contrary, our predictive framework aims to predict one target given a specific input, i.e., predicting solely ground-state geometries. To obtain the quantum property of a molecule, generative methods compute and average the properties of all generated molecular geometries. However, recent studies on 3D GNNs~\cite{schutt2017schnet, liu2021spherical, klicpera2020directional, klicpera2020fast} show that merely using the ground-state geometry with the lowest total energy can significantly boost the performance of the property prediction task. Hence, we propose a different one-to-one method to predict the unique ground-state geometries directly. Compared to generative methods, our predictive framework interprets and formulates the 3D geometry reconstruction problem from a distinct and novel perspective. 

\textbf{SSL Methods.}
A recent work~\citep{liu2021pre} proposes to pre-train regular GNNs with 3D geometry data of molecular graphs in a contrastive self-supervised fashion. Intuitively, it learns representations of 2D graphs that share the maximum mutual information with the 3D geometry data. The SSL methods and our framework are different directions to utilizing 3D geometries benefiting downstream property predictions. Moreover, SSL methods capture 3D information implicitly without the inference of atom coordinates, making it limited to molecular property prediction scenarios and cannot work for additional analyses where explicit structures are required.

\textbf{Synthetic Coordinates.}
The previous work~\citep{klicpera2021directional} is motivated by the same assumption that 3D molecular geometries are essential to many properties of molecules but is orthogonal to our work. Instead of studying the prediction of ground-state geometries, it focuses on the downstream property prediction tasks given generated 3D geometries. The synthetic coordinates of atoms are computed based on atom distance bounds from ETKDG methods provided by RDKit~\cite{rdkit}. A drawback is that the performance of property predictions is limited by the quality of geometry computed by RDKit. And the geometry computation is still computationally expensive for large-scale datasets. We later show that our method outperforms RDKit in terms of both accuracies of geometry prediction and downstream prediction, and consumes significantly less computation time.

\section{Dataset and Benchmark}
To provide a testbed for 3D molecular geometry analysis, we develop the Molecule3D benchmark. In this section, we describe details on the benchmark, including the dataset,
and evaluation protocols.

\subsection{Dataset}
The curated Molecule3D dataset provides 3,899,647 molecules with highly precise ground-state geometries optimized by DFT. We curate this dataset because it currently lacks a large-scale dataset to enable adequate learning for the proposed one-to-one geometry prediction task. Existing datasets are inapplicable to either providing non-ground-state geometries or limited ground-state geometries with constricted atom types. Therefore, we emphasize that the Molecule3D dataset is the key to making the proposed geometry predictive task feasible. Details about the Molecule3D dataset, including data statistics, data splits, data source, data processing procedure, and comparison with existing geometry datasets, are provided in Appendix~\ref{append: dataset}.

\subsection{Evaluation Protocols}
\label{sec: eval}
This section describes how to evaluate predicted 3D geometries from three aspects, including the error of distances, validity in 3D space, and benefits to the downstream prediction of Homo-Lumo gap. 

\textbf{Error of Distances.}
As aforementioned in Section~\ref{sec: framework}, in the 3D geometry prediction problem, pairwise distances of the molecule are targeted to be predicted due to their invariance to translation and rotation of the molecule. Hence, we propose to evaluate the performance of the 3D geometry prediction with the error of predicted pairwise distances. Table 1 shows that error over distances is consistent with the commonly used metric, root-mean-square deviation (RMSD) of 3D coordinates. However, computing error over distances is slightly faster than computing RMSD. In addition, when 3D coordinates are not explicitly computed, the error of pairwise distances can still be an effective metric, but RMSD cannot be obtained.
The computation of two proposed metrics, mean absolute error (\textsc{MAE}) and root-mean-square error (\textsc{RMSE}) are provided in Appendix~\ref{append: error_dist}.


\textbf{Validity of Geometries.}
Besides the error over pairwise distances, we propose to evaluate predicted geometries from another perspective, that is, validity. Specifically, we evaluate whether the model can successfully reconstruct atom coordinates in 3D space. An EDM stores squared Euclidean distances between a set of points. Given a molecule of $n$ atoms, its ground-truth EDM is denoted as $\hat{D} \in \mathbb{R}^{n \times n}$. The squared distance is given as $\hat{D}_{ij}=\|\boldsymbol{\hat{p}_i}-\boldsymbol{\hat{p}_j}\|_2^2$, where $\boldsymbol{\hat{p}_1}, ..., \boldsymbol{\hat{p}_n} \in \mathbb{R}^3$ represent the ground-truth atom positions in 3D space. The predicted distance matrix is denoted as $D \in \mathbb{R}^{n \times n}$. When we formulate to predict 3D coordinates directly, $D_{ij}=\|\boldsymbol{p_i}-\boldsymbol{p_j}\|_2^2$, where $\boldsymbol{p_1}, ..., \boldsymbol{p_n} \in \mathbb{R}^3$ represent the predicted 3D coordinates. Alternatively, we can formulate to firstly predict the distance matrix $D$, and then reconstruct 3D coordinates by $D$ based on the relationship between $D$ and its Gram matrix $M \in \mathbb{R}^{n \times n}$. In this case, the predicted distance matrix $D$ may not be a valid EDM. Even if $D$ is a valid EDM, it may reconstruct atom coordinates in space higher than 3D due to the high rank of its Gram matrix. Based on this, the third metric \textsc{Validity3D} is proposed to measure the percentage of molecules that are predicted with valid atom coordinates in 3D space:
\begin{equation}
    \textsc{Validity3D} = \frac{1}{|\mathbb{D}|} |\{D| \textrm{Rank}(M) \le 3, D \in \mathbb{D}\}|,
\end{equation}
where $M$ represents the Gram matrix of $D$. Mathematical details of reconstructing 3D coordinates based on Euclidean distance matrix are provided in Appendix~\ref{append: math_coords}.

\textbf{Error of Property Prediction.}
In addition to direct evaluation on pairwise distances and geometry validity, we propose to further evaluate the quality of predicted geometries via its performance in downstream property prediction tasks. In the second stage of the GAP framework, we use the predicted 3D geometries as fixed inputs and train a 3D GNN to predict the HOMO-LUMO gap. We evaluate the prediction performance by the mean absolute error (MAE) between the predicted properties and the ground-truth properties as shown in Appendix~\ref{append: error_hlgap}. 

\section{Experiments}
Our assessment of the proposed GAP framework with EMPNN is twofold. First, we evaluate our methods on the Molecule3D benchmark following the evaluation protocols introduced in Section~\ref{sec: eval}. Second, we evaluate the proposed methods with multiple prediction tasks in MoleculeNet, where the 3D geometry prediction is adapted as a predictive pretraining method and the pre-trained EMPNN is fine-tuned with a SchNet for downstream property predictions. The performance of our methods is then compared with other SSL pretraining methods.

\subsection{3D Geometry Prediction on Molecule3D}
\label{sec: geo_pred}
To evaluate the quality of 3D geometries predicted by EMPNN, we compare the performance with the domain knowledge-based method ETKDG provided by RDKit and two generative methods GraphDG and GeoMol. Note that the generative methods are proposed under a different problem setting that aims to learn the distribution of molecule geometries. They require molecules each with multiple non-ground-state conformers for training. Hence training generative methods on Molecule3D dataset that includes only one single ground-state conformer for each molecule may lead to model collapsing.
We therefore follow original works to train GraphDG and GeoMol on ISO17 and GEOM-QM9, respectively. Then, we evaluate their performance on the Molecule3D dataset. Detailed model configurations, training, and inference settings are provided in Appendix.

\subsubsection{Error Over Distances}
We firstly assess the accuracy of reconstructed 3D geometries by computing the error of predicted distances between atoms. In particular, we compute MAE and RMSE of the predicted intra-molecular pairwise distances to ground-truth distances. As RDKit-ETKDG can provide 3D geometries for nearly all molecules in the Molecule3D dataset, we compare its error with our EMPNN on the entire test set. However, the error of GraphDG and GeoMol cannot be obtained on the entire test set due to the slow generation procedure with a large number of failures. Therefore, we randomly sample approximately 100,000 molecules to build a test subset in which every molecule can be successfully inferred by all of these methods. Note that we do not run geometry prediction experiments with various random seeds because the Molecule3D dataset is sufficiently large to obtain relatively stable results. Besides, it is hard to repeat all experiments with various random seeds for such a large dataset.

Results in Table~\ref{tab: error_geo} show that EMPNN achieves more accurate ground-state 3D geometries than RDKit-ETKDG, GraphDG, and GeoMol in terms of MAE and RMSE over pairwise distances between atoms. We believe that the limited performance of generative methods GraphDG and GeoMol is due to the gap between their training datasets and the Molecule3D dataset, where the latter one includes more diverse molecules with a larger range of sizes. 
The number of expensively generated ground-truth conformers is considered a bottleneck of 3D geometry studies.
Given a fixed total number of ground-truth conformers, EMPNN is able to discover more molecules during training as it only requires a single conformer for each molecule, whereas generative methods are trained with fewer molecules due to the need for multiple conformers per molecule.
In addition, neither RDKit-ETKDG nor generative methods aim to reconstruct the ground-state geometry with the lowest energy at inference. 

\begin{table}[]
\caption{Assessment of MAE and RMSE over predicted intra-molecular pairwise distances, and RMSD over 3D geometries on Molecule3D dataset using random split. Percentage of valid 3D coordinates reconstructed (\textsc{Validity3D}) is computed on the test subset. For each metric, the best results are highlighted in bold numbers. \textit{Left}: Molecule3D validation \& test set. \textit{Right}: Molecule3D test subset}
\begin{center}
\begin{scriptsize}
\begin{sc}
\addtolength{\tabcolsep}{-3pt}
\parbox{.35\linewidth}{
    \begin{tabular}{lccc|ccc}
        \toprule
        & \multicolumn{3}{c|}{Validation} & \multicolumn{3}{c}{Test} \\
        \cmidrule{2-7}
        Method & MAE & RMSE & RMSD & MAE & RMSE & RMSD\\
        \midrule
        RDKit-ETKDG & 0.599 & 0.963 & 1.703 & 0.600 & 0.965 & 1.702 \\
        EMPNN & \textbf{0.379} & \textbf{0.672} & \textbf{1.471} & \textbf{0.380} & \textbf{0.677} & \textbf{1.473} \\
        \bottomrule
    \label{tab: error_geo_entire}
    \vspace{-10pt}
    \end{tabular}}
\hfill
\parbox{.46\linewidth}{
    \begin{tabular}{lcccc}
        \toprule
        Method & MAE & RMSE & RMSD & Validity3D\\
        \midrule
        GraphDG & 1.598 & 2.528 & 2.647 & 97.83\% \\
        GeoMol & 2.024 & 2.652 & 3.504 & 96.70\%\\
        RDKit-ETKDG & 0.565 & 0.942 & 1.610 & 99.76\%\\
        EMPNN & \textbf{0.402} & \textbf{0.739} & \textbf{1.445} & \textbf{100\%}\\
        \bottomrule
    \label{tab: error_geo_subset}
    \vspace{-10pt}
    \end{tabular}}
\label{tab: error_geo}
\end{sc}
\end{scriptsize}
\end{center}
\vspace{-20pt}
\end{table}

\subsubsection{3D Conformation of Molecules}
While errors over pairwise distances are effective metrics to evaluate 3D geometry prediction, it is also desired to reconstruct valid atom coordinates in 3D space for using them in more real-world scenarios where explicit conformations are needed. \textsc{Validity3D} is a metric we propose to measure the percentage of molecules that are successfully reconstructed with valid 3D coordinates. 

Results in the \textsc{Validity3D} column of Table~\ref{tab: error_geo}(b) show that EMPNN can achieve 100\% of \textsc{Validity3D} while all baseline methods fail on part of molecules. RDKit as a commonly used tool also has high \textsc{Validity3D} but still fails on challenging molecules. Moreover, we observe a lower \textsc{Validity3D} of GraphDG due to failures on computing 3D coordinates from generated distances with the EDG algorithm provided by RDKit. The \textsc{Validity3D} of GeoMol is limited by failures on disconnected fragments, issues about macrocycles, tetrahedral chiral centers, and other featurization issues. In contrast to baseline methods, EMPNN consistently predicts valid atom coordinates in 3D space without specific restrictions. In Appendix~\ref{append:conformation}, we further visualize conformations produced by EMPNN and baseline methods. 

\begin{wraptable}[9]{r}{5.2cm}
\vspace{-20pt}
\caption{Time needed to produce 3D geometries over the Molecule3D test subset with approximately 100,000 molecules. We time until coordinates are obtained.}
\label{tab: time}
\begin{center}
\begin{small}
\begin{sc}
\begin{tabular}{lc}
    \toprule
    Method & Time\\
    \midrule
    DFT & $>$20 days \\
    GraphDG & $\approx$10 hrs \\
    GeoMol & $\approx$13 hrs \\
    RDKit-ETKDG & $\approx$1 hrs \\
    EMPNN & \textbf{$\approx$3 mins} \\
    \bottomrule
\end{tabular}
\end{sc}
\end{small}
\end{center}
\vspace{-10pt}
\end{wraptable}

We further present a comparison of running time between different methods in Table~\ref{tab: time}. The results show that our methods are significantly more efficient than DFT and baseline methods. In other words, our methods can achieve favorable geometry prediction accuracy while speeding up the computations compared with physics-based methods and generative machine learning methods. It demonstrates that reconstructing ground-state 3D molecular geometries with predictive deep learning approaches is a promising solution to reduce computational costs.

\subsubsection{Homo-Lumo Gap Prediction from 3D Geometries}
\label{sec: direct_prop}

\begin{wraptable}{r}{8.2cm}
\vspace{-10pt}
\caption{MAE performance on predicting the HOMO-LUMO gap using different inputs. Experiments are conducted with random split on Molecuel3D test set. The best results are highlighted in bold numbers}
\label{tab: hlgap}
\begin{center}
\begin{small}
\begin{sc}
\addtolength{\tabcolsep}{-1pt}
\begin{tabular}{lcc}
    \toprule
    Input & Validation & Test\\
    \midrule
    2D Graph & 0.2088$\pm$0.0031 & 0.2085$\pm$0.0028 \\
    RDKit-ETKDG & 0.1701$\pm$0.0004 & 0.1713$\pm$0.0005 \\
    EMPNN & 0.1565$\pm$0.0006 & 0.1577$\pm$0.0008 \\
    Ground-truth 3D & \textbf{0.0699$\pm$0.0005} & \textbf{0.0698$\pm$0.0004} \\
    \bottomrule
\end{tabular}
\end{sc}
\end{small}
\end{center}
\vspace{-10pt}
\end{wraptable}

To further demonstrate the applicability of the predicted 3D geometries, we utilize the predicted 3D geometries to train a SchNet to predict the HOMO-LUMO gap. As described in Section~\ref{sec: splits}, the Molecule3D test set is further divided into train, validation, and test set for this direct property prediction task. Here, we use predicted atom coordinates as fixed inputs of SchNet. Since generative methods suffer from relatively slow inference and low \textsc{Validity3D}, impeding the generation of conformations on the entire test set, we compare EMPNN with RDKit only. To better assess the benefit of 3D information, we also experiment with 2D molecular graphs and ground-truth 3D coordinates using the same model architecture.

\begin{table*}[h]
\vspace{-15pt}
\footnotesize
\caption{Performance for molecular property prediction tasks (classification) on MoleculeNet dataset. ROC-AUC is reported over 3 runs with scaffold splitting. \underline{\textbf{Bold}} and \textbf{bold} numbers highlight the top-2 performance.}
\label{tab: prop_class}
\begin{center}
\begin{sc}
\addtolength{\tabcolsep}{-4.2pt}
\begin{tabular}{lcccccccc}
    \toprule
    Method & BBBP & Tox21 & ToxCast & Sider & ClinTox & MUV & HIV & Bace \\
    \midrule
    EdgePred & 64.5$\pm$3.1 & 74.5$\pm$0.4 & 60.8$\pm$0.5 & 56.7$\pm$0.1 & 55.8$\pm$6.2 & 73.3$\pm$1.6 & 75.1$\pm$0.8 & 64.6$\pm$4.7 \\
    AttrMask & 70.2$\pm$0.5 & 74.2$\pm$0.8 & 62.5$\pm$0.4 & 60.4$\pm$0.6 & 68.6$\pm$9.6 & 73.9$\pm$1.3 & 74.3$\pm$1.3 & 77.2$\pm$1.4 \\
    GPT-GNN & 64.5$\pm$1.1 & \underline{\textbf{75.3$\pm$0.5}} & 62.2$\pm$0.1 & 57.5$\pm$4.2 & 57.8$\pm$3.1 & 76.1$\pm$2.3 & 75.1$\pm$0.2 & 77.6$\pm$0.5 \\
    InfoGraph & 69.2$\pm$0.8 & 73.0$\pm$0.7 & 62.0$\pm$0.3 & 59.2$\pm$0.2 & 75.1$\pm$5.0 & 74.0$\pm$1.5 & 74.5$\pm$1.8 & 73.9$\pm$2.5 \\
    ContextPred & \underline{\textbf{71.2$\pm$0.9}} & 73.3$\pm$0.5 & \textbf{62.8$\pm$0.3} & 59.3$\pm$1.4 & 73.7$\pm$4.0 & 72.5$\pm$2.2 & 75.8$\pm$1.1 & 78.6$\pm$1.4 \\
    GraphLoG & 67.8$\pm$1.7 & 73.0$\pm$0.3 & 62.2$\pm$0.4 & 57.4$\pm$2.3 & 62.0$\pm$1.8 & 73.1$\pm$1.7 & 73.4$\pm$0.6 & 78.8$\pm$0.7 \\
    G-Contextual & 70.3$\pm$1.6 & 75.2$\pm$0.3 & 62.6$\pm$0.3 & 58.4$\pm$0.6 & 59.9$\pm$8.2 & 72.3$\pm$0.9 & 75.9$\pm$0.9 & \textbf{79.2$\pm$0.3} \\
    G-Motif & 66.4$\pm$3.4 & 73.2$\pm$0.8 & 62.6$\pm$0.5 & 60.6$\pm$1.1 & 77.8$\pm$2.0 & 73.3$\pm$2.0 & 73.8$\pm$1.4 & 73.4$\pm$4.0 \\
    GraphCL & 67.5$\pm$3.3 & 75.0$\pm$0.3 & \textbf{62.8$\pm$0.2} & 60.1$\pm$1.3 & 78.9$\pm$4.2 & \underline{\textbf{77.1$\pm$1.0}} & 75.0$\pm$0.4 & 68.7$\pm$7.8 \\
    JOAO & 66.0$\pm$0.6 & 74.4$\pm$0.7 & 62.7$\pm$0.6 & 60.7$\pm$1.0 & 66.3$\pm$3.9 & \textbf{77.0$\pm$2.2} & \textbf{76.6$\pm$0.5} & 72.9$\pm$2.0 \\
    GraphMVP & 68.5$\pm$0.2 & 74.6$\pm$0.4 & 62.7$\pm$0.1 & \textbf{62.3$\pm$1.6} & \textbf{79.0$\pm$2.5} & 75.0$\pm$1.4 & 74.8$\pm$1.4 & 76.8$\pm$1.1 \\
    \midrule
    GAP-EMPNN & \textbf{70.3$\pm$0.2} & \textbf{75.2$\pm$0.2} & \underline{\textbf{64.9$\pm$0.2}} & \underline{\textbf{64.0$\pm$0.6}} & \textbf{78.9$\pm$1.5} & 74.5$\pm$0.4 & \underline{\textbf{80.7$\pm$0.2}} & \underline{\textbf{82.7$\pm$0.1}} \\
    \bottomrule
\end{tabular}
\end{sc}
\end{center}
\vspace{-10pt}
\end{table*}

The property prediction results in terms of MAE are presented in Table~\ref{tab: hlgap}. Using ground-truth 3D geometries are confirmed to achieve the best result. It can be seen that using 3D geometries produced by both EMPNN and RDKit ETKDG improves the prediction performance compared to using 2D graphs only. Compared to RDKit, EMPNN predicts more accurate 3D geometries resulting in better performance on the downstream task of predicting the Homo-Lumo gap.

\subsection{Property Prediction on MoleculeNet Datasets}

We conduct experiments to show that the proposed GAP framework consisting of 3D geometry prediction and property prediction can also be used as an effective alternative to SSL pretraining framework. In Section~\ref{sec: geo_pred}, we have employed the EMPNN model to predict the ground-state 3D molecular geometries. In contrast to directly using the predicted 3D coordinates as experimented in Section~\ref{sec: direct_prop}, here we employ the EMPNN model that is pre-trained by the 3D geometry prediction task and fine-tune the EPMNN followed by a SchNet for property prediction tasks on MoleculeNet dataset. 

\begin{wraptable}{r}{7cm}
\vspace{-15pt}
\caption{Performance for molecular property prediction tasks (regression) on MoleculeNet dataset. RMSE is reported over 3 runs with scaffold splitting. \underline{\textbf{Bold}} numbers highlight the best performance.}
\label{tab: prop_reg}
\vskip 0.15in
\begin{center}
\begin{small}
\begin{sc}
\addtolength{\tabcolsep}{-3pt}
\begin{tabular}{lcc}
    \toprule
    Method & ESOL & Lipo\\
    \midrule
    AM & 1.112 & 0.730 \\
    CP & 1.196 & \underline{\textbf{0.702}} \\
    JOAO & 1.120 & 0.708 \\
    GraphMVP & 1.109 & 0.718 \\
    \midrule
    GAP-EMPNN & 0.997$\pm$0.027 & 0.732$\pm$0.005 \\
    \bottomrule
\end{tabular}
\end{sc}
\end{small}
\end{center}
\vspace{-15pt}
\end{wraptable}

Among all included pretraining methods, GraphMVP also utilizes 3D geometry information in the pretraining and is more similar to our setting. The comparison between GraphMVP and GAP-EMPNN aims to show the effectiveness of GAP-EMPNN on utilizing 3D geometry in downstream property predictions. For a fair comparison, we compare our framework with the primary version of GraphMVP instead of its variants combining both 3D SSL pretraining and regular 2D SSL pretraining. When 2D molecular graph-based pretraining is desired, the GAP-EMPNN can also cooperate with auxiliary 2D SSL tasks as GraphMVP variants do, but it is out of the scope of this evaluation.

The results of eight classification tasks are listed in Table~\ref{tab: prop_class} and the results of two regression tasks are listed in Table~\ref{tab: prop_reg}. We observe that the GAP framework with EMPNN achieves higher or comparable performance to existing SSL methods including GraphMVP on eight out of ten tasks, indicating that GAP-EMPNN can better capture 3D geometry information than GraphMVP and is hence more effective on accurate predictions on molecular properties.
On three datasets BBBP, MUV and, LIPO, we observe that both GraphMVP and GAP-EMPNN underperform regular SOTA self-supervised pretraining methods that only utilize 2D molecular graphs. This indicates that 3D geometric information is less essential to the prediction task.

\subsection{Ablation Studies}
We further perform an ablation study to demonstrate the effectiveness of the equilibrium message passing involving roto-invariant edge and angle representations of atom geometries. To ablate updates on edge and angle representations, we adopt the DeeperGCN-DAGNN\cite{liu2021fast} model that has the same architecture as EMPNN but updates node representations only without angle representations nor updating edge representations. To ablate angles, we adopt the regular MPNN~\cite{yang2019analyzing} as the baseline. The ablation study is conducted on the Molecule3D test set with the random split. Table~\ref{tab: ab_geo} in Appendix~\ref{append:ablation} shows the comparison on 3D geometry prediction, and Table~\ref{tab: ab_hlgap} shows the comparison on Homo-Lumo prediction with predicted atom coordinates. The results justify the effectiveness of EMPNN.

\section{Conclusions and Future Directions}

In this work, we formulated a novel problem of computing ground-state 3D molecular geometry. We presented the GAP framework and a novel equilibrium message passing neural network to perform ground-state geometry prediction which benefits downstream property prediction. We further established the first benchmark Molecule3D consisting of a large AI-ready dataset with evaluation protocols for the 3D ground-state geometry analysis. We showed that the proposed GAP and EMPNN can accurately and efficiently predict ground-state geometries and effectively utilize them to improve downstream predictions.
We demonstrate that geometry predictive methods can benefit more than geometry generative methods because the ground-state geometry is the most essential, and the Molecule3D provides more diverse training molecules. 
In addition to serving geometry predictive problem, the Molecule3D dataset can be used for more purposes. For example, 3D GNNs can be trained and tested on Molecule3D, and SSL methods can be pretrained on Molecule3D. We believe that further investigation and exploitation of the proposed geometry predictive problem and our Molecule3D dataset can advance a wide range of studies related to molecular 3D geometries.





\bibliographystyle{unsrtnat}
\bibliography{reference}

\newpage
\appendix
\section{Appendix}
\section{Overview of the GAP framework}
\label{append: overview}
\begin{figure*}[h]
    \centering
    \includegraphics[width=0.98\textwidth]{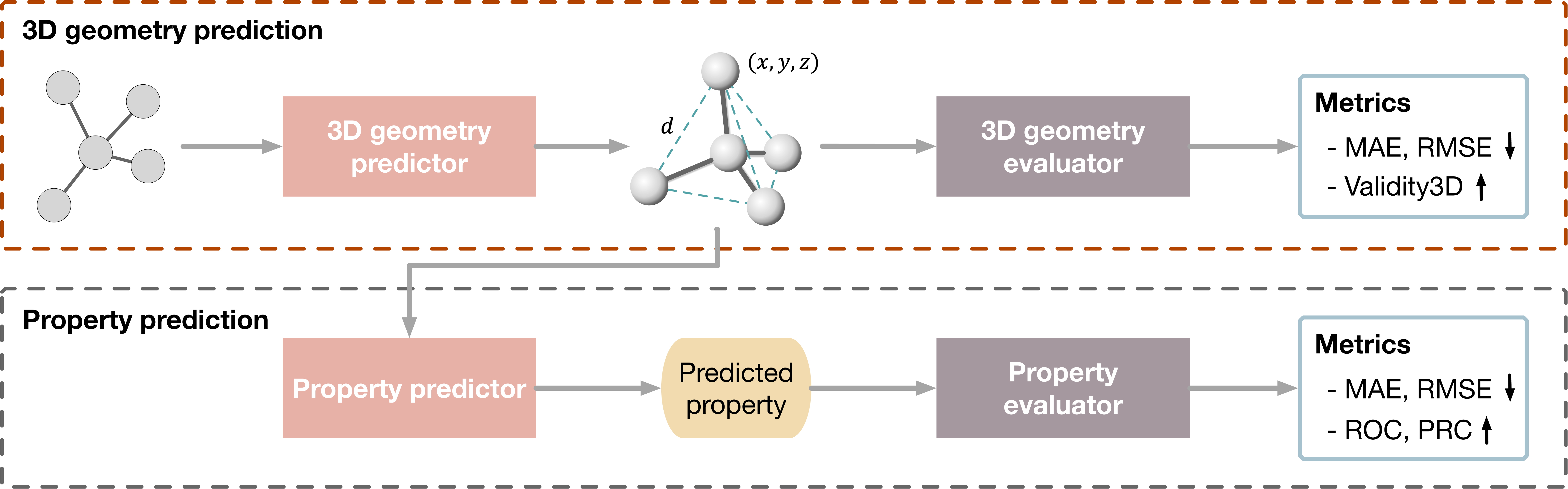}
    \caption{An overview of the Geometry Aware Prediction (GAP) framework. Given a molecular graph, we predict its ground-state geometry by employing a 3D geometry predictor. The predicted 3D geometry can be further used to benefit molecular property prediction.}
    \label{fig:framework}
\end{figure*}

An overview of the GAP framework is shown in Figure~\ref{fig:framework}.

\section{Molecule3D Dataset}
\label{append: dataset}
\subsection{Dataset Summary}
The proposed Molecuel3D dataset is curated based on the quantum chemistry database PubChemQC \cite{nakata2017pubchemqc} that provides both ground-state and excited-state 3D molecular geometries precisely computed based on DFT. However, the raw data of PubChemQC can not be easily accessed and is not ready for machine learning. From PubChemQC, we extract ground-state 3D coordinates of all valid molecules, together with four essential quantum properties, including energies of the highest occupied molecular orbital (HOMO) and the lowest unoccupied molecular orbital (LUMO), the HOMO-LUMO gap, and total energy. In addition, we provide atom features, bond features, and SMILES strings for each molecule. Finally, we encapsulate all this information in a PyTorch Geometric dataset. 

\begin{table}[h]
    \small
    \centering
    \caption{Statistics of the Molecule3D dataset.}
    \begin{tabular}{lc}
    \toprule
    $\#$Total molecules & 3,899,647 \\
    $\#$Molecules in training set& 2,339,788 \\
    $\#$Molecules in validation set& 779,929 \\
    $\#$Molecules in test set& 779,930 \\
    Splits ratio & 6:2:2 \\
    Split type & Random/Scaffold \\
    $\#$Total atoms per molecule & 29.11 \\
    $\#$Heavy atoms per molecule & 14.08 \\
    $\#$Bonds per molecule & 29.53 \\
    Included atomic types & Atomic numbers from 1 to 36 \\
    Metrics & MAE, RMSE, validity, validity3D \\
    \bottomrule
    \end{tabular}
    \label{tab:stat}
\end{table}

\subsection{Data Splits}
\label{sec: splits}
\begin{figure*}[h]
    \centering
    \includegraphics[width=0.65\textwidth]{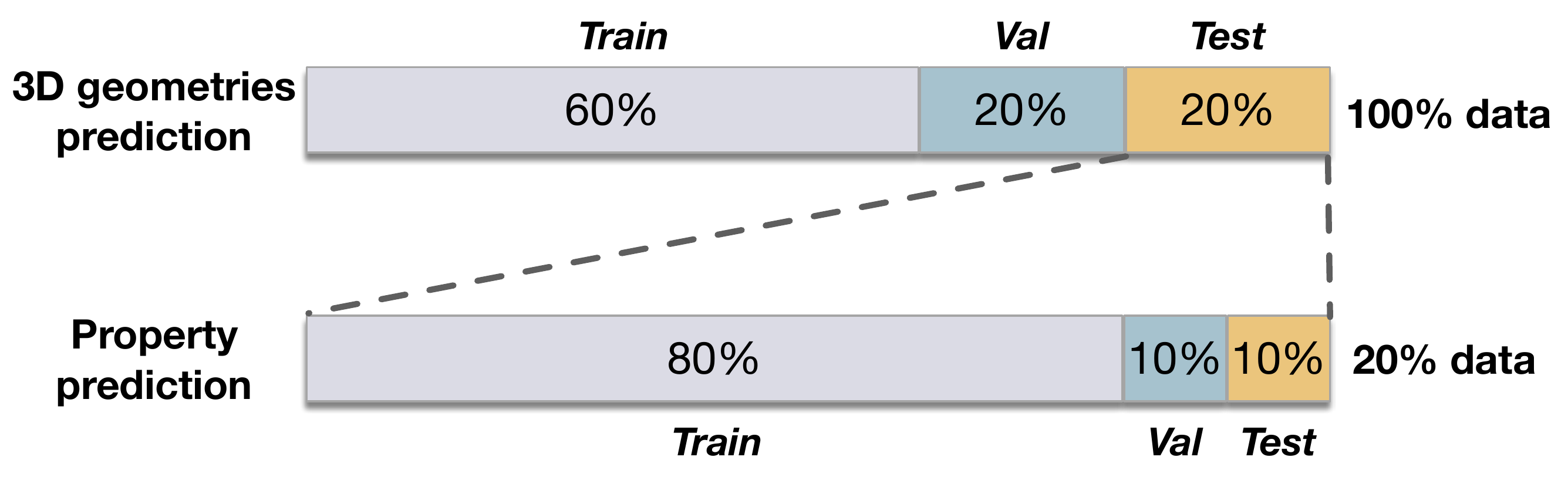}
    \caption{Data splits. For both random and scaffold split, 3D geometries prediction task involves the entire dataset, while downstream property prediction task uses only 20\% of data.}
    \label{fig:splits}
\end{figure*}

The entire Molecule3D dataset is split into $60\%/20\%/20\%$ for training/validation/test. We can train models for 3D geometry prediction on the training set and choose hyperparameters based on the validation set. Afterward, predicted 3D geometries on the test set can be used to evaluate the model performance. The 20\% test set is further split into smaller training, validation, and test sets with the ratio of 8:1:1, which is used for downstream property prediction tasks. With above described split ratio, we provide both random and scaffold splits. The illustration of data splits is shown in Figure~\ref{fig:splits}.

Together with the Molecule3D dataset, we provide two split types, random split and scaffold split. Random split ensures training, validation, and test data are sampled from the same underlying probability distribution. A molecular scaffold refers to a molecule's core component consisting of connected rings without any side chains. Splitting the dataset based on molecular scaffolds leads to a distribution shift between training and test datasets. Thus, scaffold split forces the model to capture such distribution shifts in chemical space and measures the out-of-distribution generalization ability of the model.

\subsection{Comparison with Other Datasets}
Despite their importance, there are currently only very limited benchmark datasets containing 3D molecular geometries. GEOM~\cite{axelrod2020geom} and ISO17 are two existing datasets that provide 3D molecular geometries simulated by the first principle to develop deep learning methods. They provide 400,000 and 129 molecules, respectively. Nonetheless, every molecule in GEOM and ISO17 is associated with an ensemble of multiple 3D geometries, and those geometries are unnecessary to be at the ground state. Such datasets are typically used to develop generative methods to model the distribution of 3D molecular geometries. Although the GEOM-QM9 dataset also provides ground-state geometries, it collects only 133,885 simple molecules. In QM9, heavy atoms are limited to 9 per molecule, and atomic types are restricted to H, C, N, O, and F. The QM7/QM7b and QM8 datasets from the MoleculeNet benchmark have the same constraints as QM9, and they have even fewer data than QM9. A recent benchmark Atom3D~\cite{townshend2020atom3d} collects multiple geometry datasets, but all included datasets, except for QM9, are for large compounds like proteins or RNA. Therefore, it currently lacks a large-scale dataset to enable adequate learning on the one-to-one relationship between ground-state 3D geometries and 2D molecular graphs. To make our GAP framework feasible, we curate the Molecule3D dataset with precise ground-state geometries of approximately 4 million diverse molecules derived from DFT. Although OGB-PCQM4Mv2 is curated from the same source, our Molecule3D including more molecules with 3D geometries and more molecular properties was released earlier. Besides, we provide random/scaffold splits that differ from the CID-based split in OGB-PCQM4Mv2. Moreover, with the Molecule3D, we provide evaluation protocols specifically for the proposed geometry predictive task.

Our Molecule3D is currently the largest AI-ready dataset for ground-state 3D molecular geometries. Furthermore, the Molecule3D dataset provides very diverse molecules that are actually synthesized rather than artificially created by algorithms. Therefore, the Molecule3D dataset is a challenging but practically meaningful dataset that can empower the analysis of ground-state 3D geometries.

\subsection{Source of Molecule3D}
The PubChemQC database is a large-scale public database containing 3,982,751 molecules with 3D geometries. First, PubChemQC extracts the IUPAC International Chemical Identifier (InChI)~\cite{heller2015inchi} and Simplified Molecular Input Line Entry Specification (SMILES) from its source, the PubChem database \cite{kim2016pubchem} maintained by the National Institutes of Health (NIH). Then, PubChemQC calculates both ground-state and excited-state 3D geometries of molecules using DFT at the B3LYP/6-31G* level. To expedite the calculation, the authors of PubChemQC use RICC supercomputer (Intel Xeon 5570 2.93 GHz, 1024 nodes), QUEST supercomputer (Intel Core2 L7400 1.50 GHz, 700 nodes), HOKUSAI supercomputer (Fujitsu PRIMEHPC FX100), and Oakleaf-FX supercomputer (Fujitsu PRIMEHPC FX10, SPARC64 IX 1.848 GHz). Despite these powerful computation resources, only several thousand molecules can be processed per day. Therefore, it takes years of efforts to create such an extensive database with accurate 3D molecular geometries.

\subsection{Data processing details} 
\label{sec:data_details}
The data provided by PubChemQC is not easily accessed. One needs to either use their database to query molecules one by one or download all raw data of size 2 TB. The downloaded raw data consists of millions of folders, and each contains one molecule. In every molecule folder, there are files including inp files of both ground-state and excited-state geometries and corresponding log files recording the optimization process. Note that inp files store atom coordinates only without any information about bonds. Thus, we use a mol file that is converted from the inp file by Open Babel. The mol file contains not only atom coordinates but also bonds and bond types. To make mol files more accessible by users, we read all mol files from each folder and integrate them into four SDF files. Moreover, we use the cclib package~\cite{o2008cclib} to parse log files and extract quantum properties from massive information. Extracted properties are stored in a CSV file where each column saves one property. We find some molecules invalid during the above procedure due to SMILES conversion error and RDKit warnings, with sanitize problem, or with damaged log files. Thus we remove them and ensure molecules in SDF files and CSV file are in the same order.

The raw data of our Molecule3D consists of only one folder, including four SDF files, one CSV file, and four splits files. Raw data of Molecule3D has a size of 10 GB. It is tiny compared to the original data of PubChemQC that needs 2 TB. Hence, the large dataset becomes much more accessible to users who have a concern about memory. In addition to data size reduction, we also improve the loading efficiency. By reading our raw data, users can obtain 3,899,647 molecules with ground-state geometries in seconds. Based on what we have done, other users can save at least several weeks to download, understand, parse, clean and extract molecules from the PubChemQC database. 

\section{Mathematical details for evaluation protocols}

\subsection{Error Over Distances}
\label{append: error_dist}
Given a dataset of totally $N$ intra-molecular pairwise distances between atoms, the ground-truth distance labels are denoted as $\{\hat{d_i}\}_{i=1}^N$ and the predicted distances are $\{d_i\}_{i=1}^N$, where $\hat{d}_i, d_i \in \mathbb{R}$. The first evaluation metric is mean absolute error (\textsc{MAE}) \cite{willmott2005advantages}:
\begin{equation}\label{eqn:mae}
    \textsc{MAE}\left(\{\hat{d}_i\}_{i=1}^N,\{d_i\}_{i=1}^N\right)=\frac{1}{N}\sum_{i=1}^N |\hat{d}_i-d_i|.
\end{equation}

The second evaluation metric is root-mean-square error (\textsc{RMSE}):
\begin{equation}\label{eqn:rmse}
    \textsc{RMSE}\left(\{\hat{d}_i\}_{i=1}^N,\{d_i\}_{i=1}^N\right)=\sqrt{\frac{1}{N}\sum_{i=1}^N \left(\hat{d}_i-d_i\right)^2}.
\end{equation}

\subsection{Reconstruction of 3D coordinates from EDM}
\label{append: math_coords}
According to the theorem of Gower \cite{gower1982euclidean}, the distance matrix $D$ is a valid EDM if and only if $-\frac{1}{2}JDJ$ is positive semi-definite, where $J=I-\frac{1}{n} \bm 1 \bm 1^T$ is the geometric centering matrix and $\bm 1=(1,...,1)^T \in \mathbb{R}^n$. To reconstruct atoms positions, we can leverage the relationship between $D$ and its Gram matrix $M \in \mathbb{R}^{n \times n}$, that is:
\begin{equation}
    M_{ij} = \|w_i-w_j\|_2^2 = \frac{1}{2} \left(D_{1j}+D_{i1}-D_{ij}\right).
\end{equation}
$\{\boldsymbol{w}\}_{k=1}^n$ denotes a set of direction vectors from the first atom to all atoms, and $\boldsymbol{w_k}=\boldsymbol{p_k}-\boldsymbol{p_1}$, where $\boldsymbol{p_1}, ..., \boldsymbol{p_n} \in \mathbb{R}^3$ denote predicted atom coordinates if they do exist. If $-\frac{1}{2}JDJ$ is positive semi-definite, then $M$ is also a symmetric positive semi-definite matrix and can be eigendecomposed as:
\begin{equation}
    M = U \Lambda U = \left(U \sqrt{\Lambda}\right) \left(\sqrt{\Lambda} U\right)^T = WW^T,
\end{equation} 
where $\Lambda=diag(\lambda_1,...\lambda_n)$ and $\lambda_1 \ge \lambda_2 \ge...\ge \lambda_n \ge 0$. Assuming that the first atom is placed at the origin of Cartesian space, $W$ exactly represents atom coordinates. Then, rotation and reflection can be naturally represented by applying an orthogonal matrix $Q \in \mathbb{R}^{n \times n}$. 

In short, given a molecule with its predicted distance matrix $D$, we compute the corresponding Gram matrix $M$ and then do eigendecomposition of $M$. If it yields all eigenvalues non-negative, then $D$ is a valid EDM. Otherwise, $D$ is invalid and cannot reconstruct atom coordinates in any Cartesian space. Based on this, an intermediate metric \textsc{Validity} is proposed as:
\begin{equation}
    \textsc{Validity} = \frac{1}{|\mathbb{D}|} |\{D| D \textrm{\ is\ a\ valid\ EDM}, D \in \mathbb{D} \}|,
\end{equation}
where $\mathbb{D}$ denotes the set of all predicted distance matrices. Intuitively, \textsc{Validity} measures the percentage of molecules that are predicted with valid EDM. 

Even if the predicted distance matrix is a valid EDM, it may reconstruct atom coordinates in space higher than 3D due to the high rank of the Gram matrix $M$. To reconstruct valid 3D coordinates, the predicted $D$ has to be an EDM with a corresponding $M$ of rank three or less. Specifically, all eigenvalues of $M$ are expected to be non-negative, and three or fewer of them are positive. The final metric \textsc{Validity3D} is hence proposed to measure the percentage of molecules that are predicted with valid EDM and have corresponding atom coordinates in 3D space:
\begin{equation}
    \textsc{Validity3D} = \frac{1}{|\mathbb{D}|} |\{D| \textrm{Rank}(M) \le 3, M=\textrm{Gram}(D), D \in \mathbb{D}\}|.
\end{equation}

\subsection{Error for Homo-Lumo Gap Prediction}
\label{append: error_hlgap}
Given a dataset of $m$ molecules whose HOMO-LUMO gaps are $\{\hat{y}_i\}_{i=1}^m$, and the predicted HOMO-LUMO gaps are $\{y_i\}_{i=1}^m$, where $y_i, \hat{y}_i\in\mathbb{R}$, the MAE is defined as:
\begin{equation}
	\mbox{MAE}\left(\{\hat{y}_i\}_{i=1}^m, \{y_i\}_{i=1}^m\right)=\frac{1}{m}\sum_{i=1}^{m}|\hat{y}_i-y_i|.
\end{equation}

\section{Experimental Details}
We train our models for both 3D geometries prediction and property prediction on a single 11GB GeForce RTX 2080 Ti GPU. Our experiments are implemented with PyTorch 1.8.2 and PyTorch Geometric 2.0.1, and we use RDKit 2021.09.1. 

\subsection{Model Configurations for 3D Geometry Prediction}
For the proposed EMPNN model, in addition to 5 EMP layers, we also use a virtual node that is shown to be helpful in molecular representation learning~\cite{gilmer2017neural,hu2021ogb}. Following EMP layers, we further adopt 5 propagation layers as well as an adaptive adjustment mechanism based on the DAGNN architecture~\cite{liu2020towards}. The number of hidden dimension for node, edge and angle representations is set to be 600. The output dimension is set to be 3 so that the output node representations denote atom coordinates in 3D space. The dropout rate is set to be 0. We adopt the L1 loss averaged on pairwise distances between intra-molecular atoms, and train the model with the Adam optimizer \cite{kingma2015adam} for 50 epochs. The initial learning rate is 0.0001 and decays to 80\% for every 10 epochs. We use the batch size of 20. 

\begin{table*}[h]
\caption{Summary of MoleculeNet datasets.}
\label{tab: moleculenet_summary}
\vskip 0.15in
\begin{center}
\begin{small}
\begin{sc}
\addtolength{\tabcolsep}{-3.2pt}
\begin{tabular}{lcccccccccc}
    \toprule
    & BBBP & Tox21 & ToxCast & Sider & ClinTox & MUV & HIV & Bace & ESOL & LIPO \\
    \midrule
    Tesk type & CLAS & CLAS & CLAS & CLAS & CLAS & CLAS & CLAS & CLAS & REG & REG \\
    \# Tasks & 1 & 12 & 617 & 27 & 2 & 17 & 1 & 1 & 1 & 1 \\
    \# Molecules & 2,039 & 7,831 & 8,576 & 1,427 & 1,478 & 93,087 & 41,127 & 1,513 & 1,128 & 4,200 \\
    \bottomrule
\end{tabular}
\end{sc}
\end{small}
\end{center}
\end{table*}

\begin{table*}[h]
\caption{Hyperparameters on MoleculeNet datasets.}
\label{tab: moleculenet_params}
\vskip 0.15in
\begin{center}
\begin{small}
\begin{sc}
\addtolength{\tabcolsep}{-3.8pt}
\begin{tabular}{lcccccccccc}
    \toprule
    & BBBP & Tox21 & ToxCast & Sider & ClinTox & MUV & HIV & Bace & ESOL & LIPO \\
    \midrule
    Hidden dimension & 128 & 600 & 512 & 256 & 256 & 256 & 256 & 256 & 256 & 128 \\
    \# filters & 128 & 600 & 512 & 256 & 256 & 256 & 256 & 256 & 256 & 128 \\
    Learning rate & $10^{-4}$ & $10^{-4}$ & $10^{-4}$ & $10^{-4}$ & $10^{-4}$ & $10^{-4}$ & $10^{-4}$ & $10^{-4}$ & $10^{-4}$ & $2\cdot10^{-4}$\\
    Add hydrogens & Yes & Yes & Yes & No & Yes & No & No & No & No & No \\
    \bottomrule
\end{tabular}
\end{sc}
\end{small}
\end{center}
\end{table*}

\subsection{Model Configurations for Property Prediction}
For property predictions on Molecule3D dataset and MoleculeNet datasets, we employ a SchNet composed of 6 interaction blocks. The number of gaussians is set to be 50 and the cutoff distance for interatomic interations is set to be 10\AA. For the Homo-Lumo gap prediction on Molecule3D test set, we set number of hidden dimension and number of filters to be 256, and use sum pooling to obtain graph-level representations of molecules. We employ the L1 loss and train the model with the Adam optimizer for 100 epochs. The initial learning rate is 0.0001 and decays to 96\% every 100,000 steps. The batch size is 32. For prediction tasks on MoleculeNet datasets, we choose hyperparameters as shown in Table~\ref{tab: moleculenet_params}. We consider hydrogens in geometries for some datasets but use only heavy atoms for remaining datasets. The cross entropy (CE) loss is used for classification tasks, while the L2 loss is used for regression tasks. For datasets with multiple tasks, we train a single model to learn all tasks and evaluate by the averaged performance. An Adam optimizer with the batch size of 32 is employed to train the model. To obtain graph-level representations, mean pooling is used.

\section{Visualization of the Predicted Conformations}
\begin{figure*}[h]
    \centering
    \includegraphics[width=\textwidth]{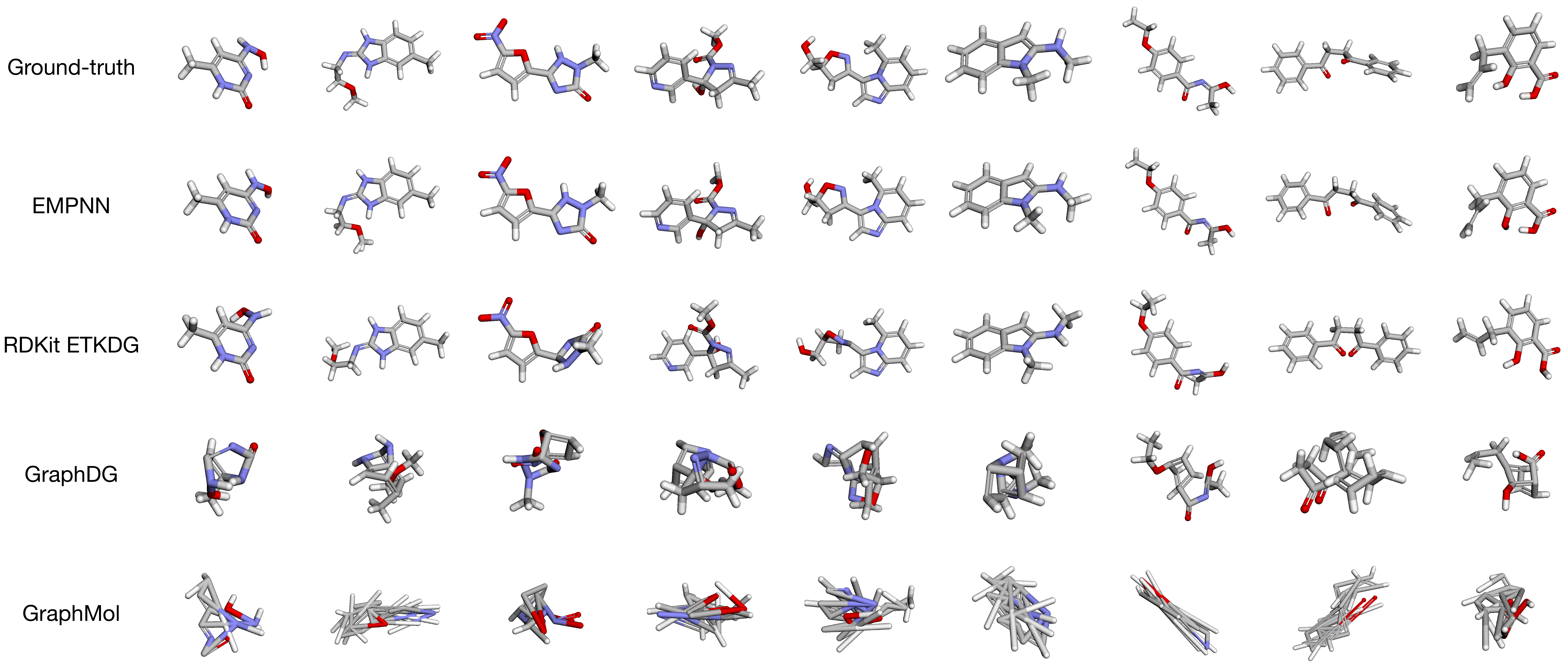}
    \caption{Conformations from the ground-truth, EMPNN and baseline methods based on molecular graphs from Molecule3D test set.}
    \label{fig:conformation}
\end{figure*}

Visualization of conformers predicted by GAP-EMPNN and baselines are shown in Figure~\ref{fig:conformation}.
\label{append:conformation}

\section{Results of Ablation Studies}
\label{append:ablation}
The results of ablation studies are shown in Table~\ref{tab: ablation}.

\begin{table}[t]
\caption{Ablation study with model architectures. \textbf{(a)} Assessment of MAE and RMSE over pairwise distances for 3D geometry prediction. \textbf{(b)} Performance of Homo-Lumo gap prediction using predicted 3D geometries. The best results are highlighted in bold numbers.}
\vskip 0.15in
\begin{center}
\begin{small}
\begin{sc}
\addtolength{\tabcolsep}{-2.5pt}
\parbox{.52\linewidth}{
    \begin{tabular}{lcc|cc}
        \toprule
        & \multicolumn{2}{c|}{Validation} & \multicolumn{2}{c}{Test} \\
        \cmidrule{2-5}
        Method & MAE & RMSE & MAE & RMSE\\
        \midrule
        edge(-)angle(-) & 0.404 & 0.711 & 0.406 & 0.716 \\
        edge(+)angle(-) & 0.401 & 0.688 & 0.402 & 0.693 \\
        edge(+)angle(+) & \textbf{0.379} & \textbf{0.672} & \textbf{0.380} & \textbf{0.677} \\
        \bottomrule
    \label{tab: ab_geo}
    \vspace{-10pt}
    \end{tabular}}
\hfill
\parbox{.45\linewidth}{
    \begin{tabular}{lcc}
        \toprule
        Method & Validation & Test \\
        \midrule
        edge(-)angle(-) & 0.1718 & 0.1734 \\
        edge(+)angle(-) & 0.1663 & 0.1676 \\
        edge(+)angle(+) & \textbf{0.1565} & \textbf{0.1577} \\
        \bottomrule
    \label{tab: ab_hlgap}
    \vspace{-15pt}
    \end{tabular}}
\label{tab: ablation}
\end{sc}
\end{small}
\end{center}
\vspace{-5pt}
\end{table}

\section{Illustration of training EMPNN}

\begin{figure*}[h]
    \centering
    \includegraphics[width=0.8\textwidth]{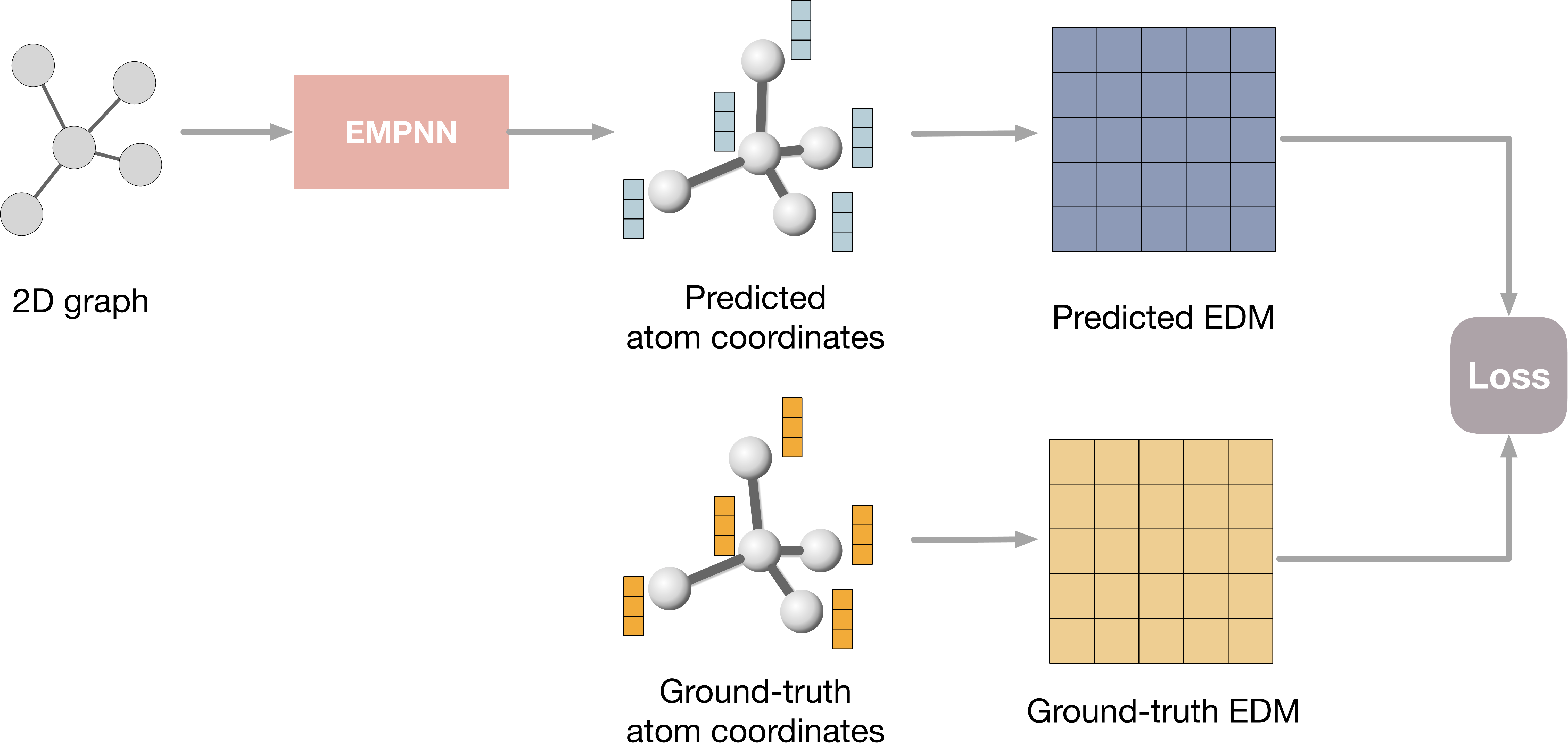}
    \caption{Illustration of the training procedure of EMPNN.}
    \label{fig:train}
\end{figure*}

The 2D graph with atom, bond, and angle features is the input of EMPNN. The EMPNN performs the equilibrium message passing as described in Section 2.2. The output of EMPNN is 3-dimensional node-level embeddings representing x, y, z coordinates in 3D space. Based on the output atom coordinates, we calculate the corresponding Euclidean distance matrix (EDM). Then, we compute L1 loss over the predicted EDM and the ground-truth EDM. Figure~\ref{fig:train} illustrates how EMPNN produces geometries and how we supervise the EMPNN on pairwise distances.

\end{document}